\documentclass[runningheads]{llncs}
\usepackage{graphicx}
\usepackage{amssymb}
\usepackage{bbm}

\usepackage[english]{babel}
\usepackage[T1]{fontenc}
\usepackage[utf8]{inputenc}

\usepackage{booktabs} 

\begin{document}
\title{Differential Diagnosis of Frontotemporal Dementia and Alzheimer's Disease using Generative Adversarial Network}
%
%

\author{ 
Da Ma\inst{1}* \and 
Donghuan Lu\inst{2,1}*\and Karteek Popuri\inst{1}
\and Mirza Faisal Beg\inst{1}**}

%
%
\institute{$^1$ Simon Fraser University, School of Engineering Science, Burnaby, V5A 1S6, Canada\\
$^1$ Tencent, Jarvis Lab, Shenzhen, 518057, China}
%


\titlerunning{Differential Diagnosis of FTD and AD using GAN}
\authorrunning{Da Ma, Donghuan Lu, Karteek Popuri, Mirza Faisal Beg}
\maketitle

%
\begin{abstract}
Frontotemporal dementia and Alzheimer's disease are two common forms of dementia and are easily misdiagnosed as each other due to their similar pattern of clinical symptoms. Differentiating between the two dementia types is crucial for determining disease-specific intervention and treatment. Recent development of Deep-learning-based approaches in the field of medical image computing are delivering some of the best performance for many binary classification tasks, although its application in differential diagnosis, such as neuroimage-based differentiation for multiple types of dementia, has not been explored. In this study, a novel framework was proposed by using the Generative Adversarial Network technique to distinguish FTD, AD and normal control subjects, using volumetric features extracted at coarse-to-fine structural scales from Magnetic Resonance Imaging scans. Experiments of 10-folds cross-validation on 1,954 images achieved high accuracy.
With the proposed framework, we have demonstrated that the combination of multi-scale structural features and synthetic data augmentation based on generative adversarial network can improve the performance of challenging tasks such as differentiating Dementia sub-types. 
\keywords{Differential Diagnosis, Magnetic Resonance Imaging, Generative Adversarial Network}
\end{abstract}

\section{Introduction}
Frontotemporal dementia (FTD) and Alzheimer's disease (AD) are two common forms of dementia sharing similar clinical symptoms, such as brain atrophy, progressive alterations in language ability, behavior and personality, and brain atrophy \cite{neary2005frontotemporal}, leading to the common misdiagnoses as each other \cite{alladi2007focal}. Although sophisticated clinical guidelines have been established for their corresponding diagnosis, such as the the NINCDS-ADRDA criteria \cite{neary1998frontotemporal}, while the specificity in the distinguishing of AD from FTD is still low even with significant expertise from clinicians ($23\%$ as reported from \cite{varma1999evaluation}). The symptomatic treatment in clinical practice varies for different dementia subtypes \cite{pasquier2005telling}. It is therefore essential to develop a computer-aided diagnosis system with improved accuracy for differential diagnosis.

Structural magnetic Resonance Imaging (MRI) is commonly used to detect pathological atrophic changes in the brain for diagnosis of neurodegenerative diseases such as AD and FTD. The atrophy patterns of these two types of dementia differs in terms of affected regions and rate of change, and the volumetric structural change measured by T1-weighted MRI is an effective biomarker for differential diagnosis \cite{davatzikos2008individual}. Example of volumetric-based features includes gray matter (GM) volume loss \cite{rabinovici2008distinct}, cortical thinning \cite{du2007different}, High-dimensional features based on GM and white matter (WM) volumes distribution of whole brain \cite{davatzikos2008individual}, as well as atrophy and shape deformity of individual structures \cite{looi2010shape}.

Most previous dementia classification studies focused on binary classification, e.g., normal control (NC) vs. FTD, NC vs. AD or FTD vs. AD, a few efforts have been attempted for direct multiclass dementia classification. Pradeep \textit{et al.} used features such as volumes, intrinsic shape, and extrinsic shape of the lateral ventricle and the hippocampus for the classification of NC, FTD and AD with PCA and multi-class support vector machine (SVM) classifier \cite{raamana2014three}. Tong \textit{et al.} has built a random undersampling boosting classifier for a five-class differential diagnostics of neurodegenerative diseases using csf, age, volume and grading features of whole brain as features \cite{tong2017five}. To the best of our knowledge, no deep learning based approach has been applied to the differential diagnosis of AD and FTD. 

In this study, a novel framework was proposed which combine multiscale structural information from MRI scans. We segmented the structural images into different patches of superpixels with each anatomical structures following a hierarchical level of patch size based on brain anatomy thereby extracting features in a coarse-to-fine manner. A multi-scale deep neural network was developed encapsulating the latent representation of features extracted at different scales, along with Generative Adversarial Network (GAN) technique for data augmentation. The proposed framework achieved $87.80\%$ accuracy through 10 fold cross validation experiment.

\section{Methods}
In the proposed framework, the original MRI images were first parcellated into patches with different level of sizes. The patch volume and local cortical thickness were extracted at each level ; 2) differential classification: a generative adversarial network was trained with the multiscale features to perform differential diagnosis among NC, AD and FTD subjects.

\subsection*{Materials}
Data used in the preparation of this article were obtained from two publicly available databases, i.e., the Alzheimer’s Disease Neuroimaging Initiative (ADNI) database and the frontotemporal lobar degeneration neuroimaging initiative (NIFD) database. The information of these two database can be found on: \url{adni.loni.usc.edu} and \url{http://memory.ucsf.edu/research/studies/nifd}, respectively.

A total of 1954 Structural MRI were used. To eliminate potential uncertanly coming from the clinical diagnosis, only images from subject with stable diagnosis during the the longitudinal follow up were included. Table \ref{table:Demog} describes the detailed demographic and clinical diagnostic information of the included subjects, including sex, age, education and clinical MMSE score. 

\begin{table}
\centering
\setlength{\tabcolsep}{0.05in}
\caption{The Demographics and clinical diagnostic information of the studied subjects. The first row shows the total number of subjects in each group, with the numbers in brackets showing the number of male and female subjects respectively. The remaining rows shows the mean and standard deviation (in brackets) about the subjects' age, education, as well as the clinical MMSE score.}
\begin{tabular}{c c c c }
\toprule
\textbf{Mean$\pm$SD} & \textbf{NC} & \textbf{AD} & \textbf{FTD} \\
\midrule
\textbf{Count (M/F)} &1063(533/530) &459(270/189) &434(266/168)\\
\textbf{Age} &72.19$\pm$8.28 &75.91$\pm$7.54 &64.69$\pm$8.51\\
\textbf{Education} &16.66$\pm$3.24 &15.13$\pm$2.58 &19.05$\pm$1.12\\
\textbf{MMSE} &29.40$\pm$1.39 &23.20$\pm$1.96 &25.36$\pm$6.12\\
\bottomrule
\end{tabular}
\label{table:Demog}
\end{table}%

\subsection*{Image Preprocessing}
Deep neural network has proven to be a powerful tool for image recognition tasks. However, a large number of labeled samples is necessary for its training, especially for high dimensional data as those used in this study ($256\times256\times256$ 3D images). The data included in this study contains a relatively large number of labeled data compared to most of other neuroimaging studies, although is still relatively small in the context of deep learning study in general. To reduce the size of network and improve the classification performance, following preprocessing steps were applied to reduce the dimension of input data by extracting preliminary features: 1) We first segmented the gray matter of each T1 structural MRI image into 87 anatomical structures (FreeSurfer 5.3 \cite{C1}). 2) Each structures were regarded as a regions of interest (ROI) and were further parcellated into smaller subregions, which we termed as "patches". A standard T1 MRI template was used to cluster voxels of each ROI into different patches through the $k$-means algorithm using spatial Euclidean distance as the clustering metrics~\cite{raamana2015thickness}. Each target MRI were then registered to the parcellated template using a symmetric non-rigid registration (LDDMM \cite{beg2005computing}). The patch-wise segmentation from the template were propagated back to the target space using the inverse deformation field. 3) Volume size and cortical thickness for each patch was extracted as features for downstream classification. The W-score ~\cite{ma2018quantitative} was calculated to remove the effect of head size and field of strength  Based on the volume size of each patch. 

Because the discriminate information among each class could exist in different scale of images, the best approach to keep as much as information as possible during feature dimension reduction is to extract features at multiple scales from coarse-to-fine manner. Therefore, the features were extracted at three different scales, i.e., 500, 1000 and 2000 voxels per patch. The number of voxels in each patch was predefined to balance between preserving the information embedded in the raw data and preventing overfitting due to oversized feature dimension comparing to the relatively small number of data samples. The resulted sub-parcellation of the structural ROIs resulted a total of 1488, 705 and 343 patches across the entire brain, respectively.

\subsection*{Generative Adversarial Network with multiscale as discriminator}
Generative Adversarial Network (GAN) \cite{goodfellow2014generative} has proven to be power tool to synthesize data, and grained great popularity in the generation of realistic natural image. Considering the limited number of available MRI scans in this study, we investigated the possibility of applying GAN to augment data samples and improve the classification performance.

\begin{figure}[htb]
\begin{minipage}[b]{1.0\linewidth}
  \centering
  \centerline{\includegraphics[width=10cm]{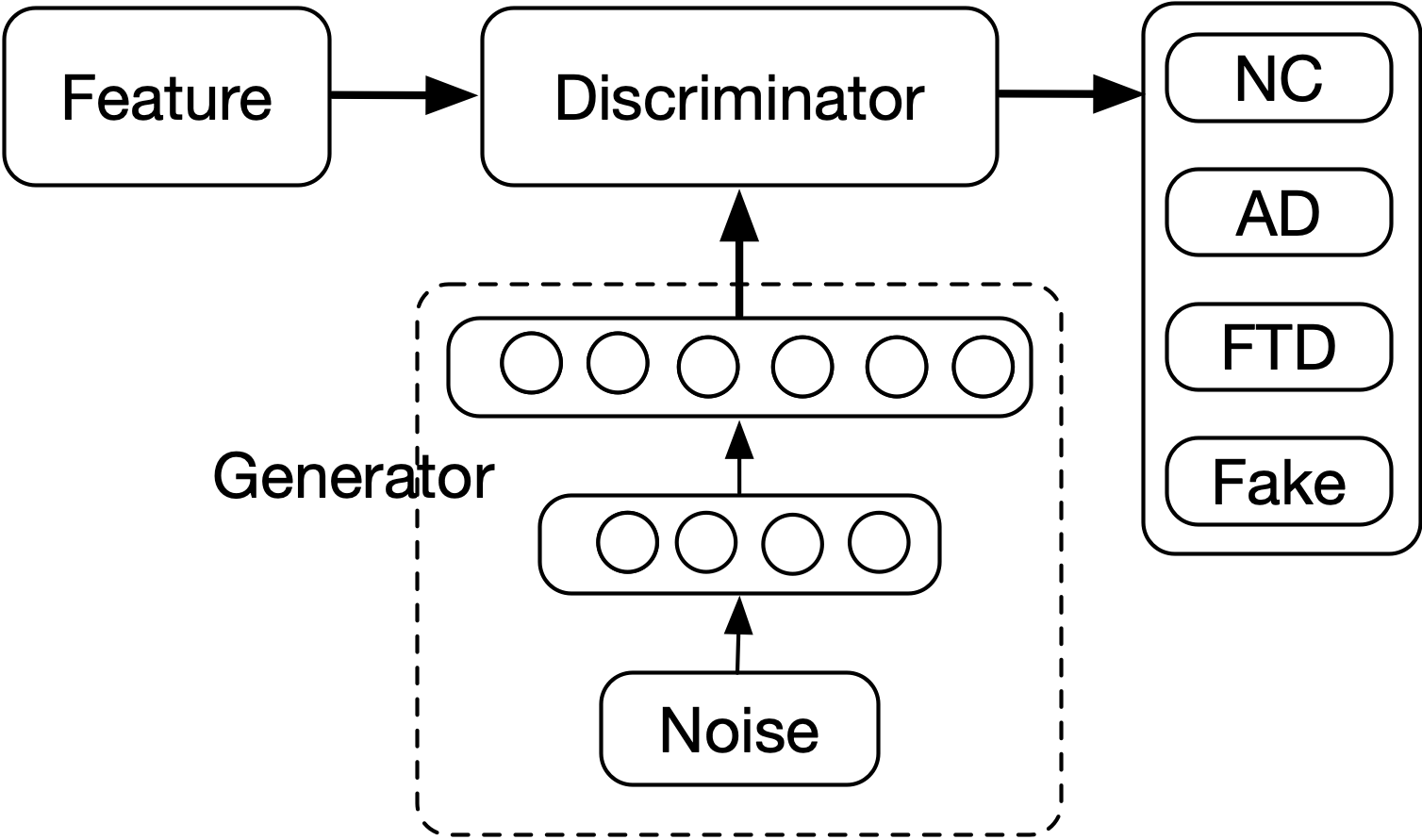}}
\end{minipage}
\caption{Architecture of Generative Adversarial Network. The number of units in generator layers are 512 and 2536(1488+705+343), respectively.}
\label{fig:Network}
\end{figure}

As shown in Figure \ref{fig:Network}, a GAN was trained with the patch-wise features extracted at multiscales from MRI. It consisted of a generator ($G$) and a discriminator ($D$). The generator aimed to find a mapping from random variables to the distribution of the actual data, while the discriminator aimed to distinguish the candidates synthesized by the generator, denoted here as "fake", from the samples belong to true classes, i.e., NC, AD and FTD. Competition between $G$ and $D$ could drive both network to improve their performance. 

The generator contained two fully connected layers with 512 and 2536 units, respectively. The number of the output unit is equal to the concatenated input size of the multiscale feature space. The unit in the hidden layer is chosen as 1/5  of the output size to contain adequate number of latent feature parameters for the generator. The activation function for the first layer was rectified linear unit (ReLU) while the one of the second layer was tanh function to squash the synthesized data into (-1,1) to ensure the symetry of the output distribution~\cite{Radford2016}.

\begin{figure}[htb]
\begin{minipage}[b]{1.0\linewidth}
  \centering
  \centerline{\includegraphics[width=10cm]{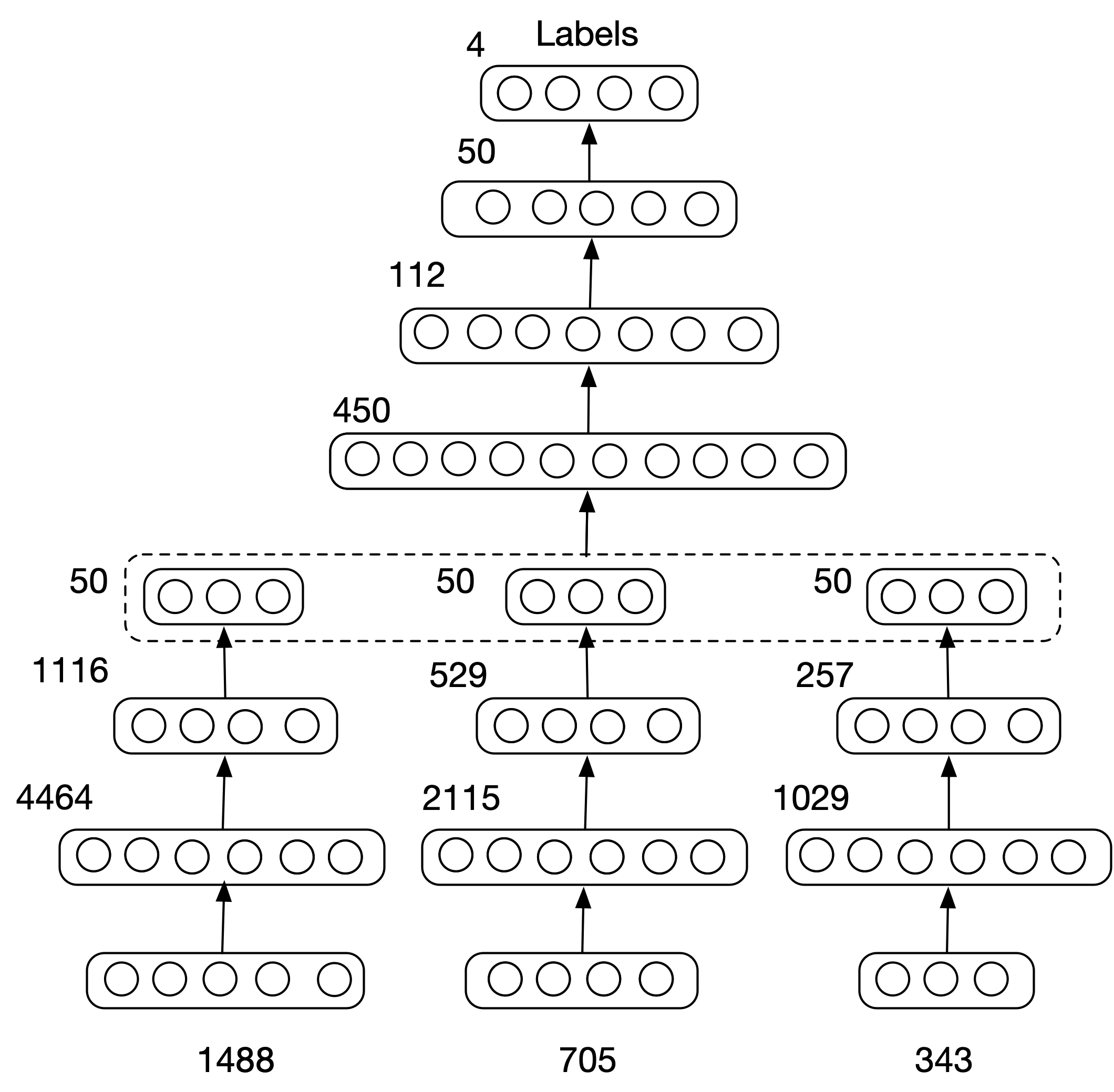}}
\end{minipage}
\caption{Architecture of the Discriminator. The number below  each layer represent the number of units. Within each MLP, if the input feature dimension was represented by $N$, the number of units for the first two hidden layers were set to $3N$ and $\frac{3}{4}N$ accordingly, with the size of the last layers all set to 50.}
\label{fig:Disriminator}
\end{figure}

A Multiscale Deep Neural Network (MDNN) \cite{lu2018multiscale} With the feature extracted at different scales was applied as the discriminator to learn the latent pattern from the feature space and perform the multi-class classification. The schematic diagram of the network were shown in Figure \ref{fig:Disriminator}. the MDNN consisted of a cascade of four four-layer Multilayer Perceptrons (MLPs). At first stage, three MLP were constructed in parallel with each using the features extracted from one of the three scale as input and were trained independently. In the second stage, the output of the previous three layers were concatenated into a single input vector which is used to feed to the fourth MLP as the input channels. The network of the entire cascaded MDNN were trained and updated at the same time. The number of units used in the whole network architecture were designed to enable the network to explore a very large range of possible latent connections among different ROI subdivisions across the whole brain at different scales, and in the mean time, while avoiding too much layers and parameters which could cause over-fitting.

\subsection*{Network Optimization}
For the optimization of GAN, the cost function was defined as:
\begin{equation}
    \min_{D}\max_{G}V(D,G)=\mathbb{E}_{x\sim p_{data}(x)}[logD(x)]+\mathbb{E}_{z\sim p_{z}(z)}[log(-D(G(z)))]
\end{equation}
where x represents input data and $p_z(z)$ is the prior on input noise variables. $log(-D(G(z)))$ was used instead of $log(1-D(G(z)))$ to avoid gradient vanish and mode collapse \cite{martin2017towards}. Root mean square propagation (RMSProp) optimizer was used to train the network and the batch size was set as 100.  

Two strategies were applied during the training process to prevent over-fitting: random dropout and early stop. Firstly, dropout layers were inserted in the network after each hidden layer of the discriminator. The dropout rate was set as 0.5 to randomly drop half of the units during training stage to prevent complex co-adaptations on training data. Secondly, during the training procedure, the optimization stops when the classification accuracy of validation set ceased to increase for 20 epochs. Furthermore, due to the limited number of data samples, early stopping with a small set of validation could still result in an unstable classification performance. Finally, an ensemble classification framework \cite{lu2018multiscale} was used. Instead of training only a single network, we randomly divide the training set further into training subset and validation subset for 10 times, resulting in 10 networks trained with independently separated split of training and validation subset. The Final classification result of testing set were obtained through majority voting of the 10 networks.

\section{Experiments and Results}

\begin{figure}[ht]
\begin{minipage}[b]{1.0\linewidth}
    \centering
    \centerline{\includegraphics[width=10cm]{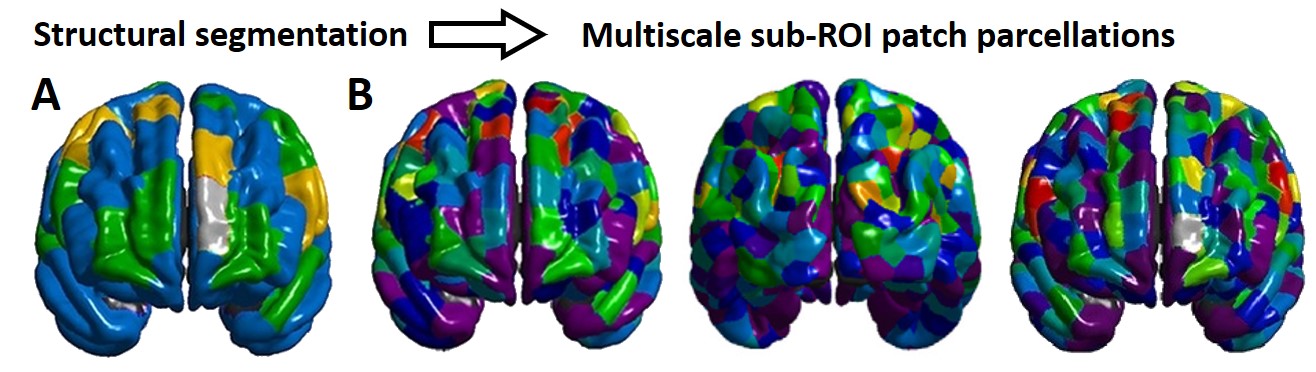}}
\end{minipage}
    \caption{Sample images of the A) initial Freesurfer structural segmentation results and B) multisale subROI patch parcellation }
    \label{fig:my_label}
\end{figure}

To validate the discriminant ability of the proposed framework, 10 fold cross validation was performed on the 1954 T1 MRI images. Because multiple images could be acquired from the same subject at different time points, the partition of the cross validation was performed on the subject-level instead of images to avoid using images from the same subject in both training and testing. Five deep learning based experiments, including MLP with single scale feature extracted at 500, 1000, 2000 voxels per patch, MDNN with multiscale features and GAN with multiscale features, were performed on those data. For benchmark comparison, we have compared the results of the MDNN with a standard classifier support vector machines (SVM) with same training/validation data splitting and cross validation setup. Principal component analysis (PCA) was used for dimension reduction on the same multiscale features and the eigenvectors representing $95\%$ of the total data variance were retained. A radial basis function (RBF) kernel was used in SVM classifier given its superior performance in classification tasks.

The accuracy and sensitivity for each class of these 6 experiments were reported in Table \ref{table:result}. The accuracy of deep learning based approaches, even with a single scale feature, were at least 3 percent higher than PCA+SVM using multiscale features. Comparing with MLP using features extracted at a single scale, the overall accuracy of MDNN using multiscale features was slightly higher, suggesting MDNN could learn hidden pattern across the small to large patches. The cross validation experiment also proved that adding the generator could further improve the classification accuracy by about $2\%$.

\begin{table}[ht]
\centering
\setlength{\tabcolsep}{0.15in}
\caption{Comparison of classification performance over different experiments. The second column is the overall classification accuracy, while the third to fifth columns represent the sensitivity of NC, AD and FTD, respectively. The first row represents the result with PCA+SVM. The second to fourth row are the classification performance of a single MLP with features extracted at different scales. The fifth row represents the MDNN result with multiscale features, and the last row represents the result of multiscale features along with data augmentation using GAN.}
\begin{tabular}{@{}|c|c|c|c|c|@{}}
\hline
           & Accuracy & NC Sen & AD Sen & FTD Sen \\ \hline
PCA+SVM    & 82.28   & 85.94 & 85.44 & 67.14  \\ \hline
500 MLP    & 85.78   & 91.60 & 82.83 & 73.31  \\ \hline
1000 MLP   & 85.41   & 90.03 & 84.91 & 73.07  \\ \hline
2000 MLP   & 85.45   & 90.34 & 82.26 & 75.06  \\ \hline
MDNN       & 85.97   & 91.05 & 83.88 & 74.20  \\ \hline
GAN        & 87.80   & 93.00 & 84.92 & 75.97  \\ \hline
\end{tabular}
\label{table:result}
\end{table}

For detailed classification result, the confusion matrix of GAN was displayed in Table \ref{table:ConfusionMatrix}. The network performances well for the task of distinguishing between AD and FTD. The discrimination between NC and FTD showed least accurate performance, leaving rooms for potential future improvement.

\begin{table}[]
\centering
\caption{Confusion matrix of GAN. The class name of the first column represent the ground truth, and the name in the first row denotes the classification result.}
\setlength{\tabcolsep}{0.2in}
\begin{tabular}{|c|c|c|c|}
\hline
    & NC  & AD  & FTD \\ \hline
NC  & 992 & 48  & 23  \\ \hline
AD  & 63  & 391 & 3   \\ \hline
FTD & 92 & 10   & 332 \\ \hline
\end{tabular}
\label{table:ConfusionMatrix}
\end{table}

\section{Conclusion and discussion}
In this study, we proposed a novel framework for accurate differential diagnosis among NC, AD and FTD pathology. The proposed methods combined the advantage of high level multi-scale feature representation learning through MDNN and the data augmentation technique using GAN. Cross validation experiments proved that, the MDNN could learn the latent representation among features extracted at different structural scales from MRI scans, and further improvement on discriminant ability can be achieved by leveraging the GAN for data augmentation. 


%
%
\bibliographystyle{splncs04}
%
\bibliography{references}
\end{document}